\documentclass[amssymb,amsmath,aps,prl,twocolumn,superscriptaddress,footinbib,a4paper]{revtex4-1}

\usepackage[dvips]{graphicx}
\usepackage[english]{babel}
\usepackage[usenames]{color}
\usepackage[normalem]{ulem}
\usepackage{amsfonts}
\usepackage{amssymb}
\usepackage{amsbsy}	
\usepackage{amsmath}
\usepackage{textcomp}
\usepackage{stmaryrd}
\usepackage[up]{subfigure}
\usepackage[colorlinks=true,linkcolor=blue]{hyperref}
\usepackage{url}
\usepackage{dcolumn}

\begin{document}

\title{Suppression of Standing Spin Waves in Low-Dimensional Ferromagnets}

\author{Andrea Taroni}
\email{andrea.taroni@physics.uu.se}
\affiliation{Department of Physics and Astronomy, Uppsala University, P.O. Box 516, 751~20 Uppsala, Sweden}
\author{Anders Bergman}
\affiliation{Department of Physics and Astronomy, Uppsala University, P.O. Box 516, 751~20 Uppsala, Sweden}
\author{Lars Bergqvist}
\affiliation{Department of Materials Science and Engineering, KTH Royal Institute of Technology, Brinellv\"{a}gen 23, 100~44 Stockholm, Sweden}
\affiliation{Department of Physics and Astronomy, Uppsala University, P.O. Box 516, 751~20 Uppsala, Sweden}
\author{Johan Hellsvik}
\affiliation{Consiglio Nazionale delle Ricerche - Superconducting and Innovative Materials and Devices (CNR-SPIN), 67100 L'Aquila, Italy}
\affiliation{Department of Physics and Astronomy, Uppsala University, P.O. Box 516, 751~20 Uppsala, Sweden}
\author{Olle Eriksson}
\affiliation{Department of Physics and Astronomy, Uppsala University, P.O. Box 516, 751~20 Uppsala, Sweden}

\date{\today}

\begin{abstract}
We examine the experimental absence of standing spin wave modes in thin magnetic films, by means of atomistic spin dynamics simulations. Using Co on Cu(001) as a model system, we demonstrate that by increasing the number of layers, the ``optical'' branches predicted from adiabatic first-principles calculations are strongly suppressed, in agreement with spin-polarized electron energy loss spectroscopy measurements reported in the literature. Our results suggest that a dynamical analysis of the Heisenberg model is sufficient in order to capture the strong damping of the standing modes.  
\end{abstract}

\maketitle

Spin-polarized electron energy loss spectroscopy (SPEELS) has recently developed to the point of becoming a powerful method with which to probe spin waves at surfaces and in thin film structures.~\cite{Plihal1999, Etzkorn2007} Landmark experiments have included the measurement of the magnon spectrum of ultrathin Co films on Cu(001),~\cite{Vollmer2003} and of a single monolayer (ML) of Fe on W(110).~\cite{Prokop2009} More recently, an asymmetry in the magnon spectrum of a bilayer of Fe on W(110) has also been reported~\cite{Zakeri2010} $-$ a direct signature of the Dzyaloshinskii-Moriya anisotropic exchange interaction present in this system.~\cite{Udvardi2009} The experimental accessibility of these properties challenge theoreticians to address a host of issues in nanoscale magnetic structures, from accurately describing the exotic ground states that arise from relativistic spin-orbit coupling effects,~\cite{Udvardi2009,Bode2007,Heide2008} to a correct treatment of the dynamics of their spin motions.

In this Letter, we address a phenomenon common to all thin ferromagnets consisting of more than one monolayer, which concerns the apparent absence of ``optical'' branches in the measured spin wave dispersion spectra.~\cite{Vollmer2003,Zakeri2010} In a slab geometry, each layer corresponds to an additional atom in the unit cell, and an elementary spin wave analysis~\cite{VanKranendonk1958} predicts the occurrence of standing wave modes.

We focus on the fcc Co/Cu(001) system, since it has been the subject of detailed experimental investigations for different Co thicknesses.~\cite{Gradmann1993,Vaz2008,Vollmer2003,Etzkorn2004,Vollmer2004b,Sinclair1960} For a system of 8 ML, Vollmer \emph{et al}.~\cite{Vollmer2003} have shown that it can be described by a nearest-neighbour Heisenberg model, characterized by a dispersion $\hbar\omega = 8JS(1-\cos(qa_0))$ along the $\langle 110 \rangle$ direction, where $J$ is the exchange coupling, $S$ is the magnitude of the spin per primitive unit cell, $q$ is the length of the magnon wave vector and $a_0 = 2.55$ \AA{}. The fit of this curve to the measured SPEELS data was $JS=15.0 \pm 0.1$ meV, which compares well with the value of $JS=14.7 \pm 1.5$ meV obtained from neutron scattering experiments on bulk fcc Co at long wavelengths.~\cite{Sinclair1960} Measurements performed on systems as thin as 2.5 MLs~\cite{Etzkorn2004,Vollmer2004b} revealed a weak thickness dependence of the magnon energies. However, for all the thicknesses reported, there was no trace of any spin wave modes other than the lowest energy acoustic branch. 

Previous theoretical investigations of Co/Cu(001) thin films, carried out using a random phase approximation to a description of the spin response of the itinerant electron system,~\cite{Costa2004a,Costa2004b} have underlined the role of particle hole excitations known as Stoner excitations. Broadly speaking, these are not relevant at low energies, where the density of magnon states dominates, but their role becomes dominant at higher energies, suppressing the magnons through a process known as Landau damping.~\cite{Mohn2003} By including them in their theoretical analysis, Costa \emph{et al.}~\cite{Costa2004a,Costa2004b} successfully described the SPEELS measurements. In particular, they correctly predict the suppression of the optical spin wave modes, and argue this occurs primarily through the Landau damping mechanism. Our results suggests that drastic damping and broadening of the magnon modes is in fact a dynamical feature of the Heisenberg Hamiltonian itself.

Our strategy is to combine first-principles calculations with Atomistic Spin Dynamics (ASD) simulations. We map an itinerant electron system onto an effective Hamiltonian with classical spins, 

\begin{equation}
  \mathcal{H} = -\frac{1}{2}\sum_{i\neq j} J_{ij} \mathbf{m}_i \cdot \mathbf{m}_j +  K \sum_{i}\left(\mathbf{m}_i \cdot \mathbf{e}_K\right) ^{2},
\label{eqn:hh}
\end{equation}

\noindent where $i$ and $j$ are atomic indices, $\mathbf{m}_i$ is the classical atomic moment, $J_{ij}$ is the strength of the exchange interaction, and $K$ is the strength of the anisotropy field along the direction $\mathbf{e}_K$. Other relativistic terms arising from spin-orbit interaction are negligible in the case of Co on Cu.

According to the Hamiltonian above, each atomic moment $\mathbf{m}_i$ experiences an effective interaction field $\mathbf{B}_i=-\frac{\partial\mathcal{H}}{\partial\mathbf{m}_i}$. The temporal evolution of the atomic spins at finite temperature is governed by Langevin dynamics, through coupled stochastic differential equations of the Landau-Lifshitz form,~\cite{Antropov1996,Skubic2008,Mentink2010}  

\begin{equation}
\frac{\partial\mathbf{m}_i}{\partial t} = - \gamma\mathbf{m}_i \times \left[\mathbf{B}_i + \mathbf{b}_i(t)\right] - \gamma\frac{\alpha}{m} \mathbf{m}_i \times \left\{ \mathbf{m}_i \times\left[\mathbf{B}_i + \mathbf{b}_i(t)\right] \right\},
\label{eqn:LLG}
\end{equation}

\noindent where $\gamma$ is the the electron gyromagnetic ratio, and $\mathbf{b}_i$ is a stochastic magnetic field with a Gaussian distribution, the magnitude of which is related to the phenomenological damping parameter $\alpha$, which eventually brings the system to thermal equilibrium. The coupled equations of motion (\ref{eqn:LLG}) can be viewed as describing the precession of each moment about an effective interaction field, with complexity arising from the fact that, since all moments are moving, the effective field is not static. 

We have calculated the exchange parameters by using the method due to Liechtenstein, Katsnelson and Gubanov,~\cite{Liechtenstein1984,Liechtenstein1986} in which the parameters are obtained from small angle perturbations from the ground state using the magnetic force theorem.~\cite{Macintosh1980} The method relies on an adiabatic approximation, in which the slow motion of the spins is decoupled from the fast motion of the itinerant electrons. We have determined the exchange values based on the relaxed geometric structure, which we calculated using the projector augmented wave method,~\cite{Kresse1999} as implemented in the Vienna \emph{ab initio} simulations package.~\cite{Kresse1996}  

Once calculated, the exchange parameters $J_{ij}$ may be Fourier transformed to obtain the so-called adiabatic magnon spectrum. In the simple case of a single monolayer (corresponding to one atom per cell), the energy of a spin wave with respect to a ferromagnetic ground state is given by $E(\mathbf{q}) = \sum_{j\neq 0} J_{0j}\left[ \exp\left( i \mathbf{q}\cdot\mathbf{R}_{0j} \right) -1 \right]$, where $\mathbf{R}_{ij}$ is the relative position vector connecting sites $i$ and $j$. From this expression it is straightforward to calculate the spin wave dispersion $\omega(\mathbf{q})$.~\cite{Kubler2009} It is important to recognise that in systems with more than one atom per cell, the spin wave energies are given by the eigenvalues of an $n\times n$ matrix, where $n$ is the number of atoms per cell.~\cite{Kubler2009} Consequently, in a thin film ferromagnet consisting of $n$ monolayers, one expects to obtain $n$ spin wave branches. 

The principal advantage of combining first-principles calculations with the ASD approach is that it allows us to address the dynamical properties of spin systems at finite temperatures.~\cite{Skubic2008,Chen1994} Two important quantities we focus on in particular are the space- and time-displaced correlation function,

\begin{equation}
  C^k (\mathbf{r}-\mathbf{r'},t) = \langle m_{\mathbf{r}^k}(t) m_{\mathbf{r'}^k}(0) \rangle - \langle m_{\mathbf{r}^k}(t) \rangle \langle m_{\mathbf{r'}^k}(0) \rangle,
  \label{eqn:cf}
\end{equation}

\noindent where the angular brackets signify an ensemble average and $k$ the cartesian component, and its Fourier Transform, the dynamical structure factor

\begin{equation}
  S^k(\mathbf{q},\omega) = \frac{1}{\sqrt{2\pi}N} \int_{-\infty}^{\infty} e^{i\omega t} \sum_{\mathbf{r},\mathbf{r'}} e^{i\mathbf{q}\cdot(\mathbf{r}-\mathbf{r'})} C^k (\mathbf{r}-\mathbf{r'},t) dt,
  \label{eqn:sf}
\end{equation}

\noindent where $\mathbf{q}$ and $\omega$ are the momentum and energy transfer, respectively. $S(\mathbf{q},\omega)$ is the quantity probed in neutron scattering experiments of bulk systems~\cite{Lovesey1984}, and can analogously be applied to SPEELS measurements. By plotting the peak positions of the structure factor along particular directions in reciprocal space, the spin wave dispersions may be obtained.~\cite{Bergman2010,Skubic2008,Chen1994} 

\begin{figure}
     \centering
     \subfigure[]{
          \label{fig:1a}
          \includegraphics[width=4.15cm]{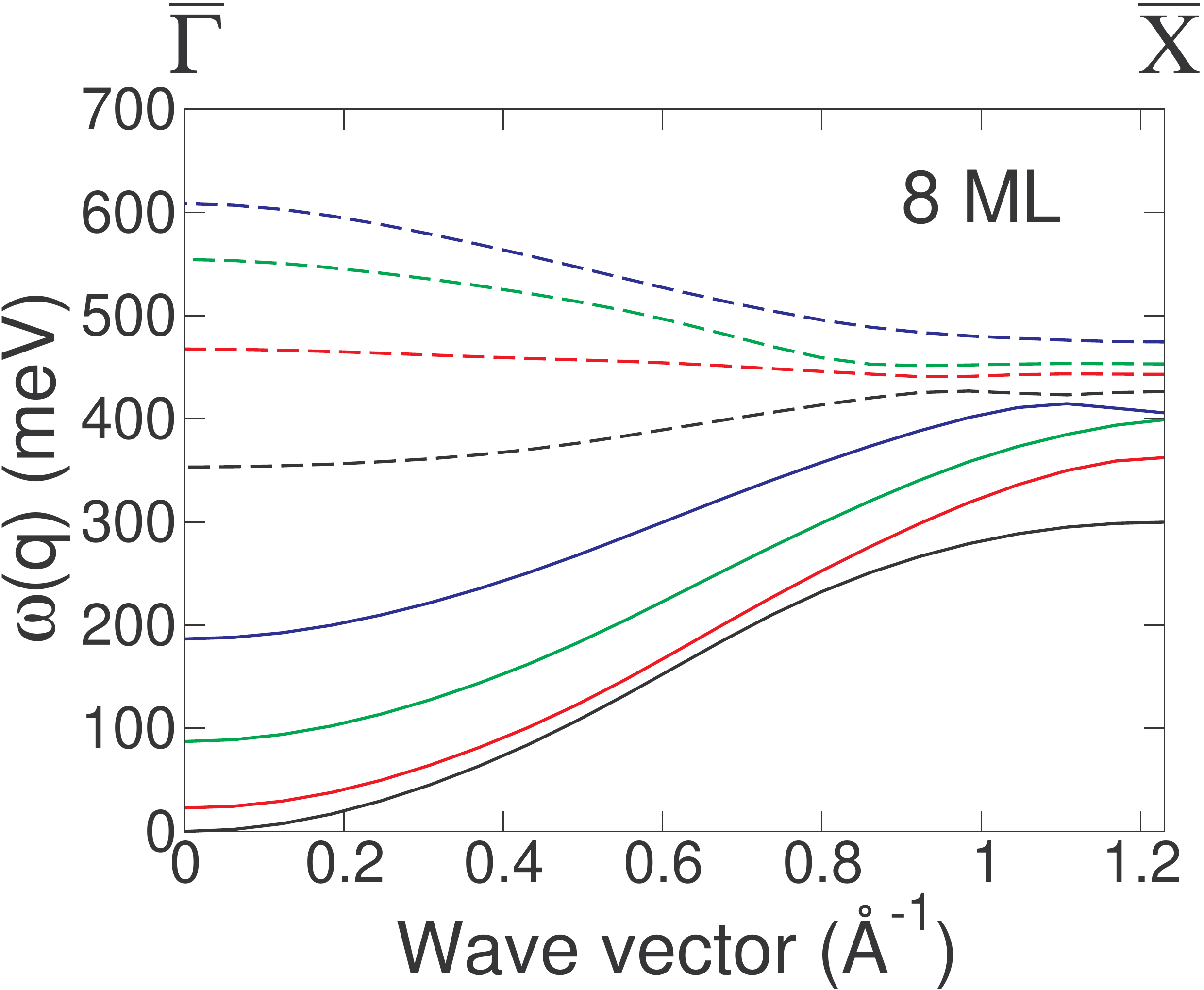}}
     \subfigure[]{
          \label{fig:1b}
          \includegraphics[width=4.15cm]{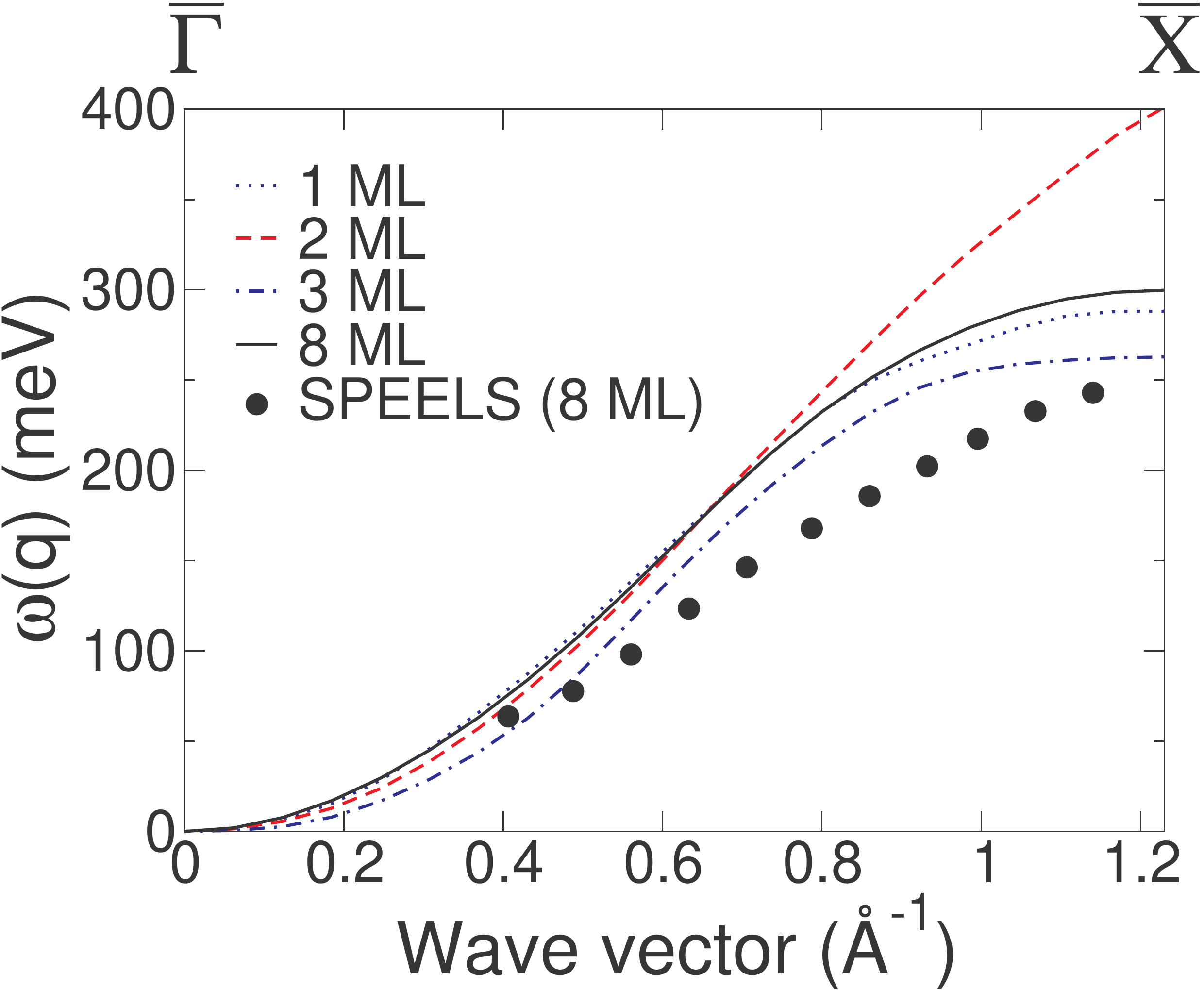}}
     \caption{\label{fig:1}(Color online) (a) Adiabatic magnon spectra for 8 ML Co/Cu(001). Note the optical branches. (b) Comparison between the acoustic branches for systems of different thickness with the SPEELS data reported for 8 ML Co on Cu(001).~\cite{Vollmer2003}}
\end{figure}

Fig. \ref{fig:1a} displays the adiabatic magnon spectrum we obtained for 8 ML Co/Cu(001). The most notable feature is the presence of several branches, one for each Co layer present. At the Brillouin zone center we observe several low energy modes. A number of these are energetically close to the acoustic branch, and na\"{i}vely one would not expect these to be washed out by Stoner excitations. However, only the lowest acoustic branch, or Goldstone mode, is detected experimentally.~\cite{Vollmer2003} As Vollmer \emph{et al.} point out,~\cite{Vollmer2003} this indicates the shortcomings of a static interpretation of their data in terms of the Heisenberg model. 

Putting these reservations aside for the time being, Fig.~\ref{fig:1b} compares the acoustic branches we have calculated for 1, 2, 3 and 8 ML films with the observed SPEELS data. By measuring the curvature of the dispersions of the lowest mode as $\mathbf{q}\rightarrow 0$, we obtain the spin wave stiffness, $D$. The comparison with experiment is reasonable: the values of $D$ we obtain are of the order of 430 meV \AA{}$^2$ - a 15 \% overestimate of the experimentally determined value of 360 meV \AA{}$^2$, but considerably softer than the theoretical value for 1 ML determined by Pajda \emph{et al.} of 532 $\pm$ 9 meV \AA{}$^2$,~\cite{Pajda2000} also obtained by a real-space adiabatic approach. We also find that the magnon energy at the $\mathrm{\overline{X}}$ point in the Brillouin zone boundary varies non-monotonically with film thickness within the range 260$-$400 meV. For 8 ML, we find a magnon energy of 300 meV, which is comparable to the experimental value of 240 meV.~\cite{Vollmer2003}

\begin{figure}
  \includegraphics[width=8.5cm]{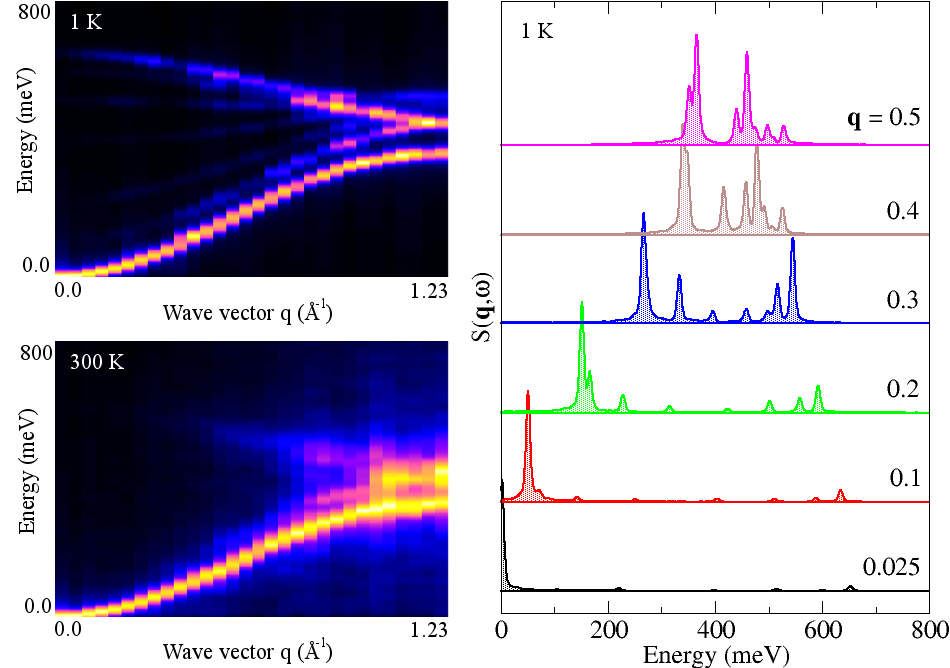}
  \caption{\label{fig:2}(Color online) Dynamical structure factor $S(\mathbf{q},\omega)$ obtained from ASD simulations of 8 ML Co/Cu(001). Top left: Magnon spectrum calculated at 1 K, with $\alpha=3\cdot10^{-4}$. Bottom left: Magnon spectrum calculated at 300 K, with $\alpha=0.05$. Right: Profile of $S(\mathbf{q},\omega)$, in which the peaks associated with the optical branches are more easily visible.}
\end{figure}

Next, we examine the dynamical behaviour by following the structure factor $S^k(\mathbf{q},\omega)$ defined by Eq. (\ref{eqn:sf}), under different conditions. The top left hand plot in Fig.~\ref{fig:2} displays the spin wave spectrum that results from a calculation performed at 1~K, in which the damping constant $\alpha$ in Eq.~(\ref{eqn:LLG}) is set to $3\cdot10^{-4}$, a value that ensures a very weak coupling to the temperature bath. Even in this limiting case, in which the role of the temperature is minimized, there is little trace of the optical branches, especially at small wave vectors. This is made clear in the right hand plot of Fig. \ref{fig:2}, in which $S(\mathbf{q},\omega)$ is plotted at selected values of $\mathbf{q}$. The bottom left-hand plot displays the spin wave spectrum we obtain at 300~K, and using a physically more plausible damping constant $\alpha=0.05$. The contrast with the spectrum directly obtained from first-principles shown in Fig.~\ref{fig:1a} is stark: for values of $q$ below 0.6, there is no trace of the optical branches. They do remain visible at shorter wavelengths, in particular when approaching the $\mathrm{\overline{X}}$ point. We also note that at 300~K the magnon energies at the zone boundary are reduced by roughly 25 meV, relative to the data calculated at 1~K, reflecting the influence of temperature effects.

\begin{figure}
  \includegraphics[width=7.5cm]{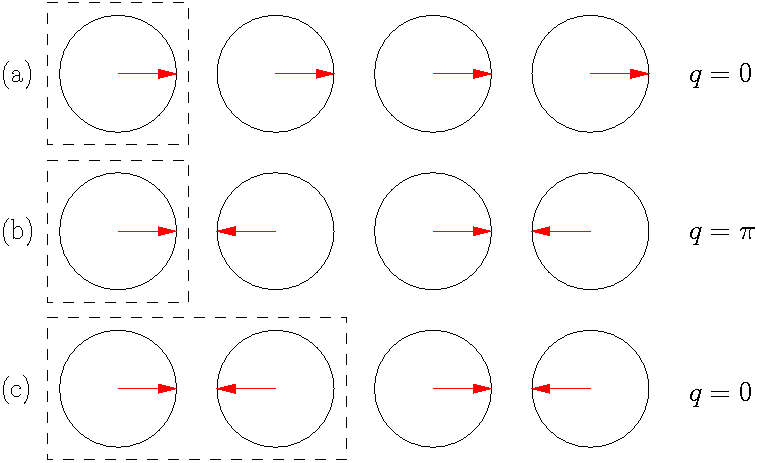}
  \caption{\label{fig:3} Magnons in an atomic chain, plotted in the $xy$ plane. (a) Long wavelength acoustic magnons; (b) acoustic magnons at the BZ boundary; and (c) the optical mode of long wavelength magnons for a system with a two atom basis. In each case the unit cell is outlined by a dashed box.}
\end{figure}

The suppression of the optical branches near the Brillouin zone center cannot be explained on thermal grounds alone, and is a reflection of the dynamical properties of the Heisenberg model. To shed light on this issue, we consider an atomic chain of spins magnetized along the $z$ axis. In this case, the magnon excitations occur only along the chain direction ($x$ direction). Fig.~\ref{fig:3} depicts the eigenmodes projected on the $xy$ plane for three separate cases: long wavelength acoustic modes, acoustic modes at the Brillouin Zone (BZ) boundary, and the optical mode at long wavelengths for a system with a two atom basis. The modes shown in Fig.~\ref{fig:3}(b) and Fig.~\ref{fig:3}(c) differ only in terms of the number of atoms in the unit cell and, just as for the Co/Cu(001) system, a calculation of the dynamical structure factor for these two modes results in a suppression of the optical mode. This effect is easily understood by considering the $x$ component of $S^k$ in Eqn.~(\ref{eqn:sf}). When $q\rightarrow 0$, the sum over cartesian coordinates for the optical mode (Fig.~\ref{fig:3}(c)) is exactly zero, since it is a sum of alternating parallel and antiparallel correlations, leading to a suppression of $S(\mathbf{q},\omega)$. Note that in the case of a one atom basis (Fig.~\ref{fig:3}(b)), the same configuration will instead result in a pronounced (and experimentally detectable) structure factor at the BZ boundary, due to the phase factor $e^{i\mathbf{q}\cdot(\mathbf{r}-\mathbf{r'})}$.

This situation is similar to that encountered for magnon excitations in rare earth systems with an hcp structure,~\cite{Jensen1991} for which it may be shown that the susceptibility is

\begin{equation}
  \overline{\overline{\chi}} ({\mathbf{q}}+{\boldsymbol{\tau}},\omega) = \frac{1}{2} \overline{\overline{\chi}}_{\mathrm{Ac}}({\bf q},\omega)(1+\cos \varphi) + \frac{1}{2} \overline{\overline{\chi}}_{\mathrm{Op}}({\bf q}, \omega)(1-\cos \varphi).
  \label{eqn:jm}
\end{equation}

\noindent For Co/Cu(001), ${\bf q}$ is a vector inside the primitive BZ, $\boldsymbol{\tau}$ is an ordering wave vector, and $\varphi={\boldsymbol{\tau}} \cdot \boldsymbol{\rho}$, where $\boldsymbol{\rho}$ is the vector coupling the two magnetic sublattices. If $\mathbf{q}$ is parallel to the $\langle 110\rangle$ direction, the phase $\varphi$ is then $h \pi + k \pi$, where $h$ and $k$ are Miller indices. Inside the primitive BZ, the phase $\varphi=0$ and, in the limit $\mathbf{q}\rightarrow 0$, Eq.~(\ref{eqn:jm}) only yields the acoustic term, and it is only this mode one expects to detect in experiment. Conversely, outside the first BZ (where $h=1$ and $k=0$) the situation is reversed and the optical mode dominates. 

\begin{figure}
  \includegraphics[width=8.5cm]{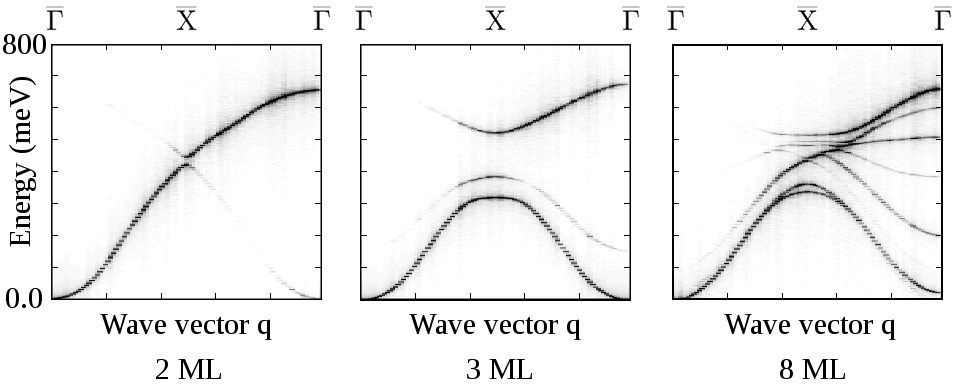}
  \caption{\label{fig:4} Dynamical structure factor $S(\mathbf{q},\omega)$ measured over the first and second Brillouin zones, for Co/Cu(001) systems consisting of 2, 3 and 8 monolayers.}
\end{figure}

To further illustrate our point, Fig.~\ref{fig:4} displays the simulated spectra of Co/Cu(001) consisting of 2, 3 and 8 MLs. In the original experiments the Co films were magnetized along the [1$\overline{1}$0] direction,~\cite{Vollmer2003} and in our simulations the precession of the moments is around this axis. The antisymmetry with respect to the $\mathrm{\overline{X}}$ point is clear, and we conclude that the suppression of the optical modes in the first BZ is a general phenomenon, in agreement with experiment. We also note that in a SPEELS experiment, the scattered electrons do not penetrate beyond the first few monolayers at the surface of the sample.~\cite{Etzkorn2007} Simulations of the structure factor for only the top layer in the 8 ML Co/Cu(001) structure, reflecting a possible dynamical measurement of only the top layers, do not differ substantially from our results calculated for the entire system.

To summarize, we have studied the experimentally observed suppression of standing spin wave modes in thin magnetic structures, through a combination of first principles calculations and atomistic spin dynamics simulations. We find that a dynamical treatment of the Heisenberg model is sufficient to capture a drastic dampening of these modes near the Brillouin zone center. Additional dissipation mechanisms, most notably in the form of Stoner excitations at higher values of $\mathbf{q}$,~\cite{Costa2004a,Costa2004b} exacerbate the suppression to the extent that in practice all trace of these modes is lost in experiment. Given the rich spin dynamics of bulk systems characterized by non-collinear ground states,~\cite{Haraldsen2010a,Haraldsen2010b} we suggest their thin film equivalents as a promising area for future study.

\begin{acknowledgments}
We gratefully acknowledge the European Research Council (ERC), the Swedish Research Council (VR), and the Knut and Alice Wallenberg Foundation for financial support.~AB and LB also acknowledge eSSENCE and SERC, respectively. Computer simulations were performed at the Swedish National Supercomputer Centre. We thank B.~Hj\"orvarsson, J. Kirschner, L.~Nordstr\"om, E.~Papaioannou and Kh. Zakeri for valuable discussions.
\end{acknowledgments}

\end{document}